\documentclass[a4paper,12pt]{article}
\usepackage{amsmath}
\usepackage{amsfonts,amssymb}
\usepackage{preprint}
\begin{document}
\numberwithin{equation}{section}
\baselineskip=14pt  
\abovedisplayskip=10pt
\belowdisplayskip=10pt
\jot=5pt
%
\def\calO{{\cal O}}
\def\calH{{\cal H}}
\def\calA{{\cal A}}
\def\boldN{{\mathbf{N}}}
\def\boldC{{\mathbf{C}}}
\def\bolda{\hbox{\boldmath$\mathit{a}$}}
\def\rvac{|\,0\,\rangle}
%
%
\hrule height0pt depth0pt
\vspace*{-12pt}
\rightline{\bf RIMS-1332}
\vspace*{50pt}
\centerline{\Huge 
Recursive Fermion System in Cuntz Algebra. I 
}
\vskip15pt
\centerline{\LARGE
--- Embeddings of Fermion Algebra into Cuntz Algebra --- }
\vskip100pt
\renewcommand{\thefootnote}{\alph{footnote})}
\centerline{\large
 Mitsuo Abe\footnote{E-mail: \abemail}
 and Katsunori Kawamura\footnote{E-mail: \kkmail}
}
\vskip10pt
\centerline{\it Research Institute for Mathematical Sciences,
Kyoto University, Kyoto 606-8502, Japan}
\vskip25pt
\vskip120pt
\centerline{\itshape\bfseries Abstract}
\vskip5pt
Embeddings of the CAR (canonical anticommutation relations) algebra of 
fermions into the Cuntz algebra $\calO_2$  (or $\calO_{2d}$ more generally)
are presented by using recursive constructions.
As a typical example, an embedding of CAR onto the $U(1)$-invariant 
subalgebra of $\calO_2$ is constructed explicitly.
Generalizing this construction to the case of $\calO_{2^p}$, 
an embedding of CAR onto the $U(1)$-invariant subalgebra of $\calO_{2^p}$ 
is obtained.
Restricting a permutation representation of the Cuntz algebra, we obtain 
the Fock representation of CAR.
We apply the results to embed the algebra of parafermions of order 
$p$ into $\calO_{2^p}$ according to the Green's ansatz.
\vfill\eject
%
%
%
\section{Introduction}
It is well understood that the $C^*$-algebra\footnote{In this paper, 
we discuss only on the $\ast$-algebraic structure and avoid considering 
the $C^*$-norm structure. } of CAR (canonical anticommutation 
relations)\cite{KR} of fermions, in which the generators $a_m$ and $a_n^*$ 
$(m,\,n=1,\,2,\,\ldots)$ satisfy
\begin{eqnarray}
&& \{ a_m, \, a_n \} = \{ a^*_m, \, a^*_n \} = 0,     \label{car-1}\\
&& \{ a_m, \, a^*_n \} = \delta_{m,n}I,               \label{car-2}
\end{eqnarray} 
is isomorphic with UHF$_2$ (i.e., the uniformly hyperfinite algebra 
of Glimm type $2^\infty$), which is defined by the $C^*$-algebra 
isomorphic with (the norm closure of) $\bigcup\limits_{n=1}^\infty
M_{2^n}$ or $\bigotimes\limits^\infty M_2$ with $M_k$ denoting 
the algebra of all $k \times k$ complex matrices.
The Cuntz algebra\cite{Cuntz} $\calO_2$ is a simple $C^*$-algebra 
generated by $s_1$ and $s_2$ satisfying
\begin{eqnarray}
&& s^*_i s_j = \delta_{i,j}\,I, \qquad i,\,j=1,\,2, \\
&& s_1 s^*_1 + s_2 s^*_2 = I.
\end{eqnarray}
We define a $U(1)$ action on $\calO_2$ by
\begin{equation}
s_i \mapsto z s_i, \quad z\in\boldC,\ \ |z|=1, \ \ i=1,\,2.
\end{equation}
Then, the subalgebra $\calO_2^{U(1)}$ consisting of $U(1)$-invariant elements 
of $\calO_2$  is a linear space generated by monomials of the form
\begin{eqnarray}
&& s_{i_1} \cdots s_{i_k} s^*_{j_k} \cdots s^*_{j_1}, \label{UHF2}
\end{eqnarray}
where $i_1,\ \ldots,\ i_k,\ j_1,\ \ldots,\ j_k = 1,\ 2$. 
From the one-to-one correspondence between the matrix element 
$e_{i_1j_1} \otimes e_{i_2j_2} \otimes \cdots 
\otimes e_{i_kj_k} \otimes I \otimes I \otimes \cdots$ and \eqref{UHF2}, 
we have an isomorphism\cite{Cuntz} between UHF$_2$ and $\calO_2^{U(1)}$. 
Therefore, we have an isomorphism between CAR and 
$\calO_2^{U(1)} \subset \calO_2$.
In other words, there exists an embedding of CAR into $\calO_2$.
Although this fact may be well-known, we could find neither its explicit 
expression nor systematic study about it in literature.
Furthermore, since the Cuntz algebra is finitely generated, its algebraic 
structures seem more manageable than that of CAR. 
Indeed, we can make many nontrivial unital $\ast$-endomorphisms in the Cuntz 
algebra explicitly as will be shown in our succeeding paper.\cite{AK2}  
Thus, if we have explicit expressions for the embedding of CAR into the Cuntz 
algebra, we can study various properties of CAR by restricting those of 
the Cuntz algebra.
\par
The parafermion algebra,\cite{Green,OK} in which the generators satisfy 
a certain set of double commutation relations, is not yet studied at all 
as a $C^*$-subalgebra of the Cuntz algebra.  
According to the Green's ansatz, we decompose each generator of the 
parafermion algebra into a sum of $p$ fermionic generators {\it commuting\/} 
with each other, where $p$ is the order of parastatistics. 
Then, the parafermion algebra of order $p$ is a subset of a tensor product of 
$p$ CAR's, which is isomorphic with 
UHF$_{2^p}\cong\calO_{2^p}^{U(1)}\subset\calO_{2^p}$.
Therefore, we may also have an embedding of a parafermion algebra into the 
Cuntz algebra.
\par
The purpose of this paper is to present the explicit expressions for the 
above embedding of the algebra of (para)fermions into the Cuntz algebra 
in a recursive way and to construct representations of the former by 
restricting those of the latter. We call a fundamental ingredient for our 
recursive construction of embeddings of CAR into $\calO_2$ (or $\calO_{2d}$ 
with $d=1,\,2,\,\ldots\,$, more generally) a {\it recursive fermion system 
(RFS).\/} It is shown that there exists a map $\zeta$ on $\calO_{2d}$ such 
that
\begin{equation}
a_n = \zeta^{n-1}(\bolda), \quad n=1,\,2,\,\ldots
\end{equation}
satisfy \eqref{car-1} and \eqref{car-2} for a suitable element 
$\bolda\in\calO_{2d}$.
In $\calO_2$, we introduce the simplest example of RFS in which the image of 
CAR is identical with $\calO_2^{U(1)}$. It should be noted, however, that 
RFS in $\calO_2$ is {\it not\/} unique, that is, an image of CAR is not 
necessarily identical with $\calO_2^{U(1)}$. We also consider a generalization 
of RFS in $\calO_{2^p}$, which we call RFS$_p$, and find that in a special 
case the image of CAR is identical with 
$\calO_{2^p}^{U(1)}\!\cong\,$UHF$_{2^p}$.
As an explicit representation of the Cuntz algebra, we consider the 
permutation representation.\cite{BJ} Restricting a certain permutation 
representation of $\calO_2$ (or $\calO_{2^p}$) to the images of CAR, we 
obtain the Fock representation of CAR. 
We apply the above results to obtain a recursive construction of the algebra 
of parafermions of order $p$ in $\calO_{2^p}$. 
We call its fundamental ingredient a {\it recursive parafermion system of 
order $p$ (RPFS$_p$).\/} The relation between RFS$_p$ and RPFS$_p$ is 
described by the Klein transformation.
What is discussed in this paper is the most fundamental aspect of the 
recursive fermion system. Importance of using the Cuntz algebra to describe 
fermion algebra will be shown in our succeeding papers.
\par
The present paper is organized as follows.  
In Sec.2, we make brief introduction of the Cuntz algebra necessary for our 
discussion in the succeeding sections. In Sec.3, the recursive fermion system 
is presented and the restriction of a permutation representation to the image 
of CAR is studied. In Sec.4, an embedding of parafermion algebra into the 
Cuntz algebra is recursively constructed by using the Green's ansatz.
The final section is devoted to discussion.
\vskip30pt
%
%
%
\section{Brief Introduction of Cuntz Algebra}
The Cuntz algebra\cite{Cuntz} $\calO_d$ $(d\geqq2)$ is a simple $C^*$-algebra 
generated by $s_1,\ s_2,\ \ldots,\ s_d$ satisfying the following relations:
\begin{eqnarray}
&& s_i^* \, s_j = \delta_{i,j} I,                     \label{CR1}\\
&& \sum_{i=1}^d s_i \, s_i^* = I,                     \label{CR2}
\end{eqnarray}
where $^*$ is a $\ast$-involution (or an adjoint operation), $I$ being the 
unit (or the identity operator). We often use the brief description such as
$s_{i_1i_2\cdots i_m}\equiv s_{i_1}s_{i_2}\cdots s_{i_m}$,
$s^*_{i_1i_2\cdots i_m}\equiv s^*_{i_m}\dots s^*_{i_2} s^*_{i_1}$
and $s_{i_1\cdots i_m;\,j_n\cdots j_1}\equiv
s_{i_1}\cdots s_{i_m} s_{j_n}^* \cdots s_{j_1}^*$.
From the relation \eqref{CR1}, $\calO_d$ is a linear space generated by 
monomials of the form $s_{i_1\cdots i_m;\,j_n\cdots j_1}$ with $m+n\geqq1$.
One should note that $\ast$-representations of $\calO_d$ are inevitably 
infinite dimensional, because \eqref{CR1} means that $s_i$'s are unitary 
in finite dimension while \eqref{CR2} prevents them from being so.
\par 
Obviously, a $d \times d$ matrix algebra $M_d$ is isomorphic with the 
subalgebra of $\calO_d$ generated by $s_{i;\,j}$. Likewise, a tensor product 
$M_d \otimes M_d$ is isomorphic with the subalgebra generated by 
$s_{i_1i_2;\,j_2j_1}$. In general, UHF$_d\cong\bigotimes\limits^\infty M_d$ 
is isomorphic with $\calO_d^{U(1)}$, which is a linear space generated by 
monomials of the form $s_{i_1 \cdots i_k;\,j_k \cdots j_1}$, $k\geqq1$.
\par
A unital $\ast$-endomorphism $\varphi$ of $\calO_d$ is defined by a mapping 
$\varphi: \calO_d \to \calO_d$ satisfying
\begin{eqnarray}
&&\varphi( \alpha X + \beta Y ) 
   = \alpha \varphi(X) + \beta \varphi(Y), \quad 
     \alpha, \, \beta \in \boldC,\ X,\,Y \in \calO_d, 
                                                      \label{unitalendo-1}\\
&&\varphi( X Y ) = \varphi(X)\varphi(Y), \quad X,\,Y \in \calO_d,      
                                                      \label{unitalendo-2}\\
&&\varphi( X^* ) = \varphi(X)^*, \quad X \in \calO_d, 
                                                      \label{unitalendo-3}\\
&&\varphi( I ) = I.                                   \label{unitalendo-4}
\end{eqnarray}
A typical example of unital $\ast$-endomorphisms of $\calO_d$ is the canonical 
endomorphism $\rho$ defined by
\begin{equation}
\rho(X) = \sum_{i=1}^d s_i X s_i^*, \quad X \in \calO_d. \label{c-endo}
\end{equation}
Indeed, from \eqref{CR1}, $\rho$ satisfies $\rho(X)\rho(Y)=\rho(XY)$.
From \eqref{CR2}, $\rho$ is unital, that is, $\rho(I)=I$, hence
$S_i\equiv\rho(s_i)$, $i=1,\,2,\,\ldots,\,d$ satisfy the relations 
\eqref{CR1} and \eqref{CR2}.
\par
In the present paper, we consider the following $\ast$-representation $\pi_s$ 
of $\calO_d$ on a countable infinite-dimensional Hilbert space $\calH$.
Let $\{e_n\}_{n=1}^\infty$ be a complete orthonormal basis of $\calH$.  
We define the action of $\pi_s(s_i)$ $(i=1,\,\ldots,\,d)$ on $\calH$ by
\begin{eqnarray}
\pi_s(s_i) e_n \equiv e_{\mu_i(n)}, \quad \mu_i(n)\equiv d(n-1)+i,  
   \quad i=1,\,\ldots,\,d,\ \ n\in\boldN.             \label{rep1}
\end{eqnarray}
By this definition, $\pi_s(s_i)$ is defined on whole $\calH$ linearly as a 
bounded operator. Then, the action of $\pi_s(s_i)^*$ on $e_n$ is determined 
by the definition of the adjoint operation. Since for any $n\in\boldN$ there 
exists a pair $\{j,\ m\}$ which satisfy $\mu_j(m)=n$,  we consider 
$\pi_s(s_i)^*$ on $e_{\mu_j(m)}$ as follows
\begin{eqnarray}
\langle \pi_s(s_i)^*\, e_{\mu_j(m)} | e_\ell \rangle 
&=& \langle e_{\mu_j(m)} | \pi_s(s_i) e_{\ell} \rangle  \nonumber\\
&=& \langle e_{\mu_j(m)} | e_{\mu_i(\ell)} \rangle      \nonumber\\
&=& \delta_{i,j}\delta_{m,\ell}                         \nonumber\\
&=& \delta_{i,j}\langle e_m | e_\ell \rangle \quad 
       \text{ for any } \ell,                         \label{rep1-2.a}\\
\noalign{\vskip10pt\noindent where 
$\langle \,\cdot\,|\,\cdot\,\rangle$ denotes the inner product on $\calH$.  
In the third equality of \eqref{rep1-2.a}, use has been made of that 
$\mu_i$'s $(i=1,\,2,\,\ldots,\,d)$ are injections and 
$\mu_i(\boldN)\bigcap\mu_j(\boldN)=\emptyset$ for each $i\not=j$ by 
\eqref{rep1}. Hence, we obtain \vskip10pt} 
\pi_s(s_i)^*\, e_{\mu_j(m)} &=& \delta_{i,j} e_m.     \label{rep1-2}
\end{eqnarray}
It is, now, straightforward to show that $\pi_s(s_i)$ and $\pi_s(s_i)^*$ 
defined by \eqref{rep1} and \eqref{rep1-2} satisfy the relations \eqref{CR1} 
and \eqref{CR2} on any $e_n$. The above representation $\pi_s$ of $\calO_d$ 
is an example of the permutation representations\cite{BJ} with a branching 
function system $\{ \mu_i\}_{i=1}^d$. One should note that $\pi_s(s_1)$ has 
eigenvalue 1 with eigenvector $e_1$. An irreducible representation which 
satisfies this property is uniquely determined up to unitary equivalence.
We call this permutation representation {\it Rep(1)\/} or the 
{\it standard representation\/} of $\calO_d$.
\vfill\eject
%
%
%
\section{Recursive Fermion System (RFS)}
In this section, we consider how to construct embeddings of CAR into $\calO_2$ 
(or $\calO_{2d}$ with $d=1,\,2,\,\ldots $) in a systematic way.
\vskip10pt
\subsection{RFS in $\calO_{2d}$} 
Let $\bolda\in\calO_{2d}$, $\zeta:\calO_{2d}\to\calO_{2d}$ be a linear 
mapping, and $\varphi$ a unital $\ast$-endomorphism on $\calO_{2d}$, 
respectively.
A triad $R=(\bolda, \, \zeta, \, \varphi)$ is called 
a {\it recursive fermion system (RFS)\/} in $\calO_{2d}$, if it satisfies 
the following conditions:\footnote{It is possible to define a RFS satisfying 
the conditions similar to \eqref{RFS1}--\eqref{RFS3} in any unital 
$C^*$-algebra apart from its existence.}
\begin{alignat}{3}
&\phantom{ii}\mbox{i) seed condition:}
&\ \ &\bolda^2 = 0, \quad 
      \{ \bolda, \, \bolda^* \} = I,                  \label{RFS1}\\
&\phantom{i}\mbox{ii) recursive condition:}
&\ \ & \{ \bolda, \, \zeta(X) \} = 0, \ \ \zeta(X)^*=\zeta(X^*), \quad  
    X \in \calO_{2d},                                 \label{RFS2}\\
&\mbox{iii) normalization condition:}
&\ \ &\zeta(X)\zeta(Y)=\varphi(XY), \quad 
    X,\,Y \in \calO_{2d}.                             \label{RFS3}
\end{alignat}  
Here, $\bolda$ and $\zeta$ are called the {\it seed\/} and the 
{\it recursive map,\/} respectively, of the RFS $R$.  
For a given RFS $R=(\bolda, \, \zeta, \, \varphi)$, the embedding associated 
with $R$,
\begin{equation}
\varPhi_R : \mbox{CAR} \hookrightarrow \calO_{2d},
\end{equation}
is defined by specifying images of generators $a_n$ $(n=1,\,2,\,\ldots)$ 
of CAR as follows
\begin{equation}
\varPhi_R(a_n)
\equiv \zeta^{n-1}(\bolda)
\equiv (\,\underbrace{\zeta\circ\zeta\circ\cdots\circ\zeta}_{n-1}\,)(\bolda),
\quad n=1,\,2,\,\ldots.                               \label{RFS4}
\end{equation}
Indeed, from \eqref{RFS1}--\eqref{RFS4}, we obtain
\begin{eqnarray}
&&
\{ \varPhi_R(a_m), \, \varPhi_R(a_n) \} 
= \varphi^{m-1}(\{ \bolda, \, \zeta^{n-m}(\bolda) \}) 
= \varphi^{m-1}(0) = 0, \quad m\leqq n, \\
&&
\{ \varPhi_R(a_m), \, \varPhi_R(a_n)^* \} 
= \varphi^{m-1}(\{ \bolda, \, \varphi^{n-m}(\bolda^*)\})
= \varphi^{m-1}(0)=0, \quad m < n, \\
&&
\{ \varPhi_R(a_n), \, \varPhi_R(a_n)^* \} 
= \varphi^{n-1}(\{ \bolda, \, \bolda^*\}) =\varphi^{n-1}(I)=I.
\end{eqnarray}
We denote $\calA_R \equiv \varPhi_R(\mbox{CAR})\subset\calO_{2d}$.
$\calA_R$ is called the CAR subalgebra associated with $R$.
\par
A typical example of RFS in $\calO_2$ is given by the {\it standard RFS\/} 
$SR=(\bolda,\,\zeta,\,\varphi)$, which is defined by
\begin{eqnarray}
&& \bolda\equiv s_1 s_2^*,                            \label{RFS5}\\
&& \zeta(X)\equiv s_1 X s_1^* - s_2 X s_2^*, \quad X \in \calO_2, 
                                                      \label{RFS6}\\
&& \varphi(X)\equiv\rho(X)= s_1 X s_1^* + s_2 X s_2^*, \quad 
       X \in \calO_2,                                 \label{RFS7}
\end{eqnarray}
where $\rho$ is the canonical endomorphism \eqref{c-endo} of $\calO_2$.
Indeed, it is easy to show that \eqref{RFS5}--\eqref{RFS7} satisfy 
\eqref{RFS1}--\eqref{RFS3}.
For the standard RFS $SR$, we denote $\calA_S\equiv\varPhi_{SR}(\mbox{CAR})$.
Then, we have $\calA_S=\calO_2^{U(1)}$ as is easily proved by mathematical 
induction: It is obvious that $\calA_S\subset\calO_2^{U(1)}$, while 
$\calA_S\supset\calO_2^{U(1)}$ is also satisfied since any 
$s_{i_1\cdots i_k;\,j_k\cdots j_1} \in \calO_2^{U(1)}$, $k\geqq1$ is 
expressed in terms of $\varPhi_{SR}(a_n) \ (n\leqq k)$.
\par
The standard RFS $SR$ in $\calO_2$ can be easily extrapolated to the 
corresponding one $R=(\bolda,\,\zeta,\,\varphi)$ in 
$\calO_{2d}$ $(d=1,\,2,\,\ldots)$ as follows:
\begin{eqnarray}
&& \bolda\equiv \sum_{k=1}^d \epsilon_k\,s_{i_k}s_{j_k}^*, 
                                                      \label{gRFS1}\\[-5pt]
&& \zeta(X)\equiv \sum_{k=1}^{d} \epsilon_k'\,
          (s_{i_k} X s_{i_k}^* - s_{j_k} X s_{j_k}^*), \quad X \in \calO_{2d}, 
                                                      \label{gRFS2}\\[-5pt]
&& \varphi(X)\equiv\rho(X)=\sum_{i=1}^{2d} s_i X s_i^*, \quad 
       X \in \calO_{2d},
                                                      \label{gRFS3}\\[-3pt]
&& \varPhi_R(a_n) \equiv \zeta^{n-1}(\bolda), 
                                                      \label{gRFS4}
\end{eqnarray}
where an arbitrary division of indices $\{1,\,\ldots,\,2d\}$ into two 
ordered parts $\{i_1\equiv1,\,i_2,\,\ldots,\,i_d\}$ and 
$\{j_1,\,j_2,\,\ldots,\,j_d\}$ is introduced in \eqref{gRFS1};
$\epsilon_1=\epsilon_1'=+1$, $\epsilon_k,\,\epsilon_k'=\pm1$ $(k\geqq2)$.
The subalgebra of $\calO_{2d}$ generated by $\varPhi_R(a_n)$ $(n\leqq k)$ 
is isomorphic with $\bigotimes\limits^k(I_d\otimes M_2)\cong
\bigotimes\limits^k M_2$, where $I_d$ is the $d \times d$ unit matrix.
Thus, $\calA_R$ determined by \eqref{gRFS1}--\eqref{gRFS4} is isomorphic 
with UHF$_2$.
\vskip8pt
\subsection{Generalization of RFS}
As an example of RFS in $\calO_4$, we may set
\begin{eqnarray}
&& \bolda = s_1 s_2^* + s_3 s_4^*,                    \label{O_4-a}\\
&& \zeta(X) = 
      s_1 X s_1^* - s_2 X s_2^* \pm (s_3 X s_3^* - s_4 X s_4^*).
                                                      \label{O_4-phi+-}
\end{eqnarray} 
It is interesting to consider whether there are nontrivial elements 
in the form of \eqref{gRFS1} which anticommute with $\bolda$ defined by 
\eqref{O_4-a}.
We find such an element is uniquely given by
\begin{equation}
\tilde{\bolda}=s_1 s_3^* - s_2 s_4^*.                 \label{O_4-a2}
\end{equation}
Furthermore, if we require that $\zeta(X)$ anticommute with $\tilde{\bolda}$ 
as well as with $\bolda$, $\zeta$ is uniquely determined as
\begin{equation}
\zeta(X) = s_1 X s_1^* - s_2 X s_2^* - s_3 X s_3^* + s_4 X s_4^*. 
                                                      \label{O_4-phi}
\end{equation}
Thus, we can extend the RFS $R=(\bolda, \, \zeta, \, \varphi)$ in $\calO_4$ 
defined by \eqref{O_4-a}, \eqref{O_4-phi} and \eqref{gRFS3} with $d=2$ to the 
tetrad $SR_2=(\bolda_1\equiv \bolda,\,\bolda_2\equiv\tilde{\bolda};\ 
 \zeta,\,\varphi)$ in such a way that 
\begin{equation}
\begin{array}{c}
 \varPhi_{SR_2} : \mbox{CAR} \hookrightarrow \calO_4,\\[8pt]
 \varPhi_{SR_2}(a_{2(n-1)+i})
 \equiv \zeta^{n-1}(\bolda_i), \quad i=1,\,2, \ n=1,\,2,\,\ldots
\end{array}                                           \label{UHF-RFS2}
\end{equation}
gives an embedding of CAR into $\calO_4$.
In contrast with the ordinary RFS given by a triad, we call this kind of 
tetrad $R_2=(\bolda_1,\,\bolda_2;\,\zeta,\,\varphi)$ RFS$_2$.
Although we have defined a RFS$_2$ with specifying the explicit expressions 
for $R_2=SR_2$ in the above, RFS$_2$ itself can be, of course, defined 
abstractly without such a specific expression. We call the above RFS$_2$ 
denoted by $SR_2$ the {\it standard RFS\/}$_2$ since it is the special one 
in which $\calA_{S_2}\equiv\varPhi_{SR_2}(\hbox{CAR})= \calO_4^{U(1)}$: 
It is obvious that $\calA_{S_2}\subset\calO_4^{U(1)}$, while 
$\calA_{S_2}\supset\calO_4^{U(1)}$ is also satisfied since any 
$s_{i_1\cdots i_k;\,j_k\cdots j_1}\in\calO_4$, $k\geqq1$ is expressed in terms 
of $\varPhi_{SR_2}(a_n)$ $(n\leqq2k)$.
\par
Likewise, in $\calO_{2^p}$, we can generalize RFS to RFS$_{p}$ 
$R_p=(\bolda_1,\,\bolda_2,\,\ldots,\,\bolda_p,\,\zeta,\,\varphi)$ in such a 
way that 
\begin{equation}
\begin{array}{c}
\varPhi_{R_p} : \mbox{CAR} \hookrightarrow \calO_{2^p},\\[10pt]
\varPhi_{R_p}(a_{p(n-1)+i})\equiv\zeta^{n-1}(\bolda_i), 
 \quad  i=1,\,\ldots,p,\ n=1,\,2,\,\ldots
\end{array}                                           \label{RFS_p-embed}
\end{equation}
gives an embedding of CAR into $\calO_{2^p}$ and none of 
$\bolda_i\ (i=1,\,\ldots,\,p)$ is expressed as $\zeta(X)$ with 
$X\in\calO_{2^p}$. Here $\zeta$ and $\varphi$ are a linear mapping and a 
unital endomorphism on $\calO_{2^p}$, respectively;
$(\bolda_1,\,\bolda_2,\,\ldots,\,\bolda_p;\,\zeta,\,\varphi)$ should 
satisfy
\begin{alignat}{3}
&\phantom{ii}\mbox{i) seed condition: }
&\ \ &\{\bolda_i,\,\bolda_j\}=0,\quad 
     \{\bolda_i,\,\bolda_j^*\}=\delta_{i,j}I, \\
&\phantom{i}\mbox{ii) recursive condition: }
&&\{\bolda_i,\,\zeta(X)\}=0,\quad \zeta(X)^*=\zeta(X^*),\ \ 
    X \in \calO_{2^p},                                \label{RFS_p-a-phi}\\
&\mbox{iii) normalization condition: }
&&\zeta(X)\zeta(Y)=\varphi(XY),\ \ 
   X,\,Y \in \calO_{2^p}. 
\end{alignat}
The {\it standard RFS\/}$_{p}$ 
$SR_p=(\bolda_1,\,\bolda_2,\,\ldots,\,\bolda_p;\,\zeta,\,\varphi)$, 
in which $\calA_{S_p}\equiv\varPhi_{SR_p}(\hbox{CAR})=
\calO_{2^p}^{U(1)}$, is given by
\begin{eqnarray}
&& \bolda_i = \sum_{k=1}^{2^{p-i}}\sum_{\ell=1}^{2^{i-1}}
         (-1)^{\sum\limits_{m=1}^{i-1}\left[\frac{\ell-1}{2^{m-1}}\right]}
         s_{2^i(k-1)+\ell}s_{2^{i-1}(2k-1)+\ell}^*, 
   \quad i=1,\,\ldots,\,p,                            \label{RFSp-a}\\[-3pt]
&& \zeta(X) = \sum_{i=1}^{2^p}
             (-1)^{\sum\limits_{m=1}^{p}\left[\frac{i-1}{2^{m-1}}\right]}
             s_i X s_i^*,                             \label{RFSp-phi}\\[-7pt]
&& \varphi(X) = \rho(X) \equiv \sum_{i=1}^{2^p} s_i X s_i^*,
\end{eqnarray}
where $[x]$ denotes the largest integer not greater than $x$. We have 
constructed \eqref{RFSp-a} in the {\it bootstrap\/} way: For any $\bolda_i$, 
each term $s_k s_\ell^*$ in $\bolda_i$ requires that $\bolda_j$ $(j\not=i)$ 
should involve either $s_{m_j}s_k^* \mp s_{n_j}s_\ell^*$ or 
$s_k s_{m_j}^* \mp s_\ell s_{n_j}^*$ with some $m_j$ and $n_j$, and then 
$\bolda_i$ is conversely required to involve $\pm s_{m_j}s_{n_j}^*$ 
for all $j$, so that $\{\bolda_i \,|\,i=1,2,\ldots,p\}$ satisfy 
the $p$-dimensional canonical anticommutation relations. In this way, 
for a given $\bolda_1\equiv\sum\limits_{k=1}^{2^{p-1}}s_{2k-1}s_{2k}^*$, 
the other $\bolda_i$'s are uniquely constructed with the normalization 
that the sign factor for the term $s_1 s_k^*$ in $\bolda_i$ is plus. 
As for $\zeta$ in \eqref{RFSp-phi}, it is uniquely determined from 
\eqref{RFS_p-a-phi} and from the normalization that the sign factor for 
$s_1 X s_1^*$ is plus.
\vskip10pt
\subsection{Representation of RFS}
In this subsection and the next, we consider the representation of CAR, 
which is obtained by restricting the standard representation $\pi_s$ on 
$\calH$ defined by \eqref{rep1} of $\calO_2$ (or $\calO_{2^p}$) to the 
standard RFS (or the standard RFS$_p$):
\begin{equation}
\mbox{CAR} 
   \ \ \mathop{\hookrightarrow}^{\varPhi_{SR}} \ \ \calO_2 
   \xrightarrow{\ \ \ \pi_s\ \ \ } {\cal L}(\calH).
\end{equation}
Hereafter, we identify $s_i$ and $a_n$ with $\pi_s(s_i)$ and 
$(\pi_s\circ\varPhi_{SR})(a_n)$ in $\calO_2$ 
(or $(\pi_s\circ\varPhi_{SR_p})(a_n)$ in $\calO_{2^p}$), respectively,  
for simplicity of description.
\par
For the standard representation Rep(1) of $\calO_2$ defined by \eqref{rep1}, 
we find that $e_1$ is a vacuum with respect to the annihilation operators 
$a_n\ (n=1,\,2,\,\ldots)$ defined by \eqref{RFS4} with \eqref{RFS5} and 
\eqref{RFS6}, that is, it satisfies
\begin{equation}
a_n e_1 = 0 \ \text{ for } n=1,\,2,\,\ldots.          \label{rep1vac}
\end{equation}
The proof of \eqref{rep1vac} is the following: From \eqref{rep1} and 
\eqref{rep1-2}, we obtain
\begin{equation}
\begin{array}{lclclcl}
s_1^* e_{2n-1} &=& e_n, &\quad &s_1^* e_{2n} &=& 0, \\[3pt]
s_2^* e_{2n-1} &=& 0,   &\quad &s_2^* e_{2n} &=& e_n.
\end{array}
\quad n=1,\,2,\,\ldots.                               \label{rep1s*}
\end{equation}
Hence, $e_1$ is the simultaneous eigenvector both for $s_1^*$ 
with eigenvalue 1 and for $s_2^*$ with 0.
Thus, from \eqref{RFS4} with \eqref{RFS5} and \eqref{RFS6}, we have
\begin{equation}
a_n\,e_1 = \zeta^{n-1}(\bolda)\,e_1 
= s_1^{n-1}\,\bolda\, e_1
=0.                                                   \label{RFSvacproof}
\end{equation}
The (antisymmetric) Fock space with the vacuum $e_1$ is generated by 
operation of monomials $a_{n_1}^*\cdots a_{n_k}^*$ on $e_1$. 
Since $a_m^*$ and $a_n^*$ anticommute with each other, we have only to 
consider the case $n_1<\cdots<n_k$. 
For $k=1$, from \eqref{rep1} and \eqref{rep1s*}, we have
\begin{eqnarray}
a_{n_1}^*\,e_1 
&=& \zeta^{n_1-1}(\bolda^*)\,e_1 \nonumber\\[-2pt]
&=& s_1^{n_1-1}\bolda^*\,e_1     \nonumber\\[-2pt]
&=& s_1^{n_1-1}s_2\,e_1          \nonumber\\[-2pt]
&=& e_{N(n_1)}, \qquad
N(n_1)\equiv 2^{n_1-1} + 1.                           \label{1-particle}
\end{eqnarray}
For $k=2$, from \eqref{CR1}, \eqref{rep1},  \eqref{RFS4}, \eqref{RFS5}, 
\eqref{RFS6} and \eqref{1-particle}, we have
\begin{eqnarray}
a_{n_1}^* a_{n_2}^* \, e_1 
&=& \zeta^{n_1-1}(\bolda^*)\,s_1^{n_2-1}\,s_2 \, e_1   \nonumber\\[-2pt]
&=& s_1^{n_1-1}\,\bolda^*\,s_1^{n_2-n_1}\,s_2 \, e_1   \nonumber\\[-2pt]
&=& s_1^{n_1-1}\,s_2\,s_1^{n_2-n_1-1}\,s_2 \, e_1      \nonumber\\[-2pt]
&=& e_{N(n_1,n_2)}, \qquad
N(n_1,n_2)\equiv 2^{n_1-1}+2^{n_2-1}+1.
\end{eqnarray}
Likewise, in general, we obtain
\begin{eqnarray}
a_{n_1}^* a_{n_2}^* \cdots a_{n_k}^* \, e_1 
&=& s_1^{n_1-1}\,s_2\,s_1^{n_2-n_1-1}\,s_2\,
    s_1^{n_3-n_2-1}\,s_2\,\cdots\,
    s_1^{n_k-n_{k-1}-1}\,s_2 \, e_1                    \nonumber\\[-4pt]
&=&e_{N(n_1,\cdots,n_k)}, \label{RFSfock}\\ 
N(n_1,\cdots,n_k) &\equiv& 2^{n_1-1} + 2^{n_2-1} 
  + \cdots + 2^{n_k-1} + 1.                           \label{binary}
\end{eqnarray}
Since it is obvious that any $n\in\boldN$ is expressible in the form of 
$N(n_1,\cdots,n_k)-1$ (binary expression), any $e_n$ is uniquely given 
in the form of \eqref{RFSfock}, that is, $e_1$ is a cyclic vector in 
$\calH$ and there is no vacuum annihilated by $a_n$ $(n=1,\,2,\,\ldots)$ 
other than $e_1$. Therefore, the restriction of Rep(1) of $\calO_2$ to 
$\calA_S$ is the Fock representation with the unique vacuum.
\par
One should note that any vector $e_N$ instead of $e_1$ of the basis in Rep(1) 
of $\calO_2$ can be set as the vacuum of Fock representation by the Bogoliubov 
transformation exchanging {\it finite\/} number of the annihilation/creation 
operators such as
\begin{eqnarray}
\noalign{\vskip-3pt}
&&
a_n \mapsto 
  a_n'  
    = \begin{cases}    
        a_n^* & \mbox{for $n \in \{n_1,\,n_2,\,\ldots,\, n_k\}$,}\\[3pt]
        a_n   & \mbox{otherwise,}
      \end{cases}
\end{eqnarray}
where the set of indices $\{n_1,\,n_2,\,\ldots,\, n_k\}$ corresponds to $N$ 
through the relation \eqref{binary}.
\vskip10pt
\subsection{Representation of RFS$_p$}
As for the restriction of Rep(1) of $\calO_{2^p}$ to the standard RFS$_p$, 
it is straightforward to generalize the previous results.
From \eqref{rep1} and \eqref{rep1-2} with $d=2^p$, we have
\begin{equation}
 s_i^* e_{2^p(n-1)+j} = \delta_{i,j}e_n, \quad 
   i,\,j = 1,\,2,\,\ldots,\,2^p,\,  n=1,\,2,\,\ldots,
\end{equation}
hence, for \eqref{RFS_p-embed} with \eqref{RFSp-a} and \eqref{RFSp-phi}, 
we obtain
\begin{equation}
 a_{p(m-1)+i}\, e_1 = s_1^{m-1}\bolda_i\,e_1 = 0, \quad 
   i=1,\,2,\,\ldots,p,\, m=1,\,2,\,\ldots.
\end{equation}
Thus, $e_1$ of Rep(1) of $\calO_p$ is a vacuum for the annihilation 
operators $a_n\ (n=1,\,2\,\ldots)$ of the standard RFS$_p$, and the 
corresponding Fock space is generated by 
$a_{n_1}^*a_{n_2}^*\cdots a_{n_k}^*\,e_1$ with
$1\!\leqq\! n_1\!<\!n_2\!<\!\cdots\!<n_k$, $k\!\geqq\!1$. 
Their explicit expressions are obtained in the following.
\par 
In case $1\leqq m_1<m_2<\cdots<m_k$ for $n_j=p(m_j-1)+i_j$, 
$i_j=1, \ldots,p$, $j=1,\ldots,k$, likewise in \eqref{RFSfock}, we have
\begin{eqnarray}
&&a_{p(m_1-1)+i_1}^*a_{p(m_2-1)+i_2}^*\cdots a_{p(m_k-1)+i_k}^* \, e_1
 \nonumber\\
&&\qquad\qquad
 = s_1^{m_1-1} s_{\mbox{\tiny$\!M$}_{\!1}+1} 
    s_1^{m_2-m_1-1} s_{\mbox{\tiny$\!M$}_{\!2}+1} 
    \cdots
    s_1^{m_k-m_{k-1}-1} s_{\mbox{\tiny$\!M$}_{\!k}+1} 
    \, e_1, \label{rfsp-rep1-1} \\
&& M_j \equiv 2^{i_j-1}, \quad j=1,\,2,\,\ldots,\,k.
\end{eqnarray}
On the other hand, in case $(m_{j-1}<\,)\,m_j=m_{j+1}=\cdots=m_{j+\ell}\,
(\,<m_{j+\ell+1})$ and $i_j<i_{j+1}<\cdots<i_{j+\ell}$ with some $j$ 
and $\ell$ for $n_j=p(m_j-1)+i_j$, $i_j=1, \ldots,p$, $j=1,\ldots,k$, 
the corresponding factor 
$s_1^{m_j-m_{j-1}-1} s_{_{\!M_j+1}}
\cdots s_1^{m_{j+\ell}-m_{j+\ell-1}-1} s_{_{\!M_{j+\ell}+1}}$
in rhs of \eqref{rfsp-rep1-1} is replaced by
\begin{eqnarray}
\noalign{\vskip-10pt}
&&s_1^{m_j-m_{j-1}-1} s_{\mbox{\tiny$\!M$}_{\!j,\ell}+1}, \quad
M_{j,\ell}\equiv \sum_{i=j}^{j+\ell} M_i. \\[-15pt]
&&\nonumber
\end{eqnarray}
Therefore, using the formula
\begin{eqnarray}
&&s_1^{n_1-1}s_{i_1+1} s_1^{n_2-n_1-1}s_{i_2+1} \cdots 
     s_1^{n_k-n_{k-1}-1}s_{i_k+1}\,e_1
   = e_{N(n_1,\,i_1;\,\ldots;\,n_k,\,i_k)},   
      \qquad\quad\\[-2pt]
&&N(n_1,\,i_1;\,\ldots;\,n_k,\,i_k)\equiv 
     \sum_{j=1}^k i_j \, 2^{p(n_j-1)} +1              \label{RFSp_N}
\end{eqnarray}
for $1\!\leqq\! n_1\!<\!\cdots\!<\!n_k$ and $i_1,\ldots,i_k\!=\!1,2,\ldots, 
2^p\!-\!1$, it is straightforward to obtain
\begin{eqnarray}
\noalign{\vskip-3pt}
&&
a_{p(m_1-1)+i_1}^*\cdots a_{p(m_k-1)+i_k}^*\,e_1
= e_{N(m_1,i_1;\,\cdots;\,m_k,i_k)},  \\[-5pt]
&&
N(m_1,i_1;\,\cdots;\,m_k,i_k)
  \equiv\sum_{j=1}^k 2^{p(m_j-1)+i_j-1} + 1, \\[-15pt]
&&\nonumber
\end{eqnarray}
for $p(m_1-1)+i_1<\cdots < p(m_k-1)+i_k$, hence,
\begin{eqnarray}
&&
a_{n_1}^*\cdots a_{n_k}^*\,e_1= e_{N(n_1,\,\cdots,n_k)},  
   \quad n_1<\cdots<n_k, \\[-5pt]
&&
N(n_1,\,\cdots,n_k)\equiv\sum_{j=1}^k 2^{n_j-1} + 1,
\end{eqnarray}
which is exactly the same as \eqref{RFSfock} with \eqref{binary}.
Thus, in the same way as the standard RFS, the restriction of Rep(1) of 
$\calO_{2^p}$ to $\calA_{S_p}$ is the Fock representation with 
the unique vacuum. It should be noted that this Fock representation is 
{\it strictly common to all the standard RFS$_p$.} 
This result may be understood by the fact that the standard RFS$_p$ can be 
reduced to the standard RFS through a certain embedding of $\calO_{2^p}$ 
into $\calO_2$.\cite{AK2} 
\vskip20pt
%
%
%
\section{Recursive Parafermion System}
In this section, we apply RFS to embed the algebra of parafermions into 
the Cuntz algebra.
According to the Green's ansatz, the algebra of parafermions of order $p$ is 
described in terms of $p$ fermion algebras in which any element in one 
fermion algebra {\it commute\/} with those in others. Hence, we can embed 
the algebra of parafermions into $\calO_{2^p}$ by generalizing RFS$_p$ 
so that they satisfy the suitable properties.
\vskip10pt 
\subsection{Parafermion Algebra}
To begin with, we summarize the property of the algebra of 
parafermions.\cite{Green,OK}
\par
Let the {\it parafermion algebra of order $p\,(=1,\,2,\,\ldots)$ PFA$_p$\/} 
be the $\ast$-algebra generated by $a_n$ $(n=1,\,2,\,\ldots)$ satisfying
\begin{eqnarray}
\noalign{\vspace*{-2pt}}
&& [ a_\ell,\, [ a_m, \, a_n ] \, ] = 0,              \label{PF1}\\ 
&& [ a_\ell,\, [ a_m^*, \, a_n ] \, ] = 2\delta_{\ell,m}a_n, 
                                                      \label{PF2}\\
&& \prod_{k=0}^{p}\Big(N_n + \big(k-\frac{p}{2}\big) I \Big) = 0, 
  \quad N_n \equiv \frac{1}{2}[ a_n^*, \, a_n ],      \label{PF3}
\end{eqnarray}
where we have omitted other relations obtained from \eqref{PF1} and 
\eqref{PF2} by using the $\ast$-involution (or the adjoint operation) 
and the Jacobi identity. The case $p=1$, PFA$_1$, is nothing but CAR.
Here, we have explicitly introduced the order $p$ dependence of the algebra 
in \eqref{PF3}. In the conventional standpoint in studying parastatistical 
algebras, starting with the double (anti)commutation relations such as 
\eqref{PF1} and \eqref{PF2} only, one introduces the positive integer $p$ to 
specify the vacuum as in \eqref{pfvac2} below in considering its 
representation. Then, the additional algebraic relation describing spectrum 
of the number operator $N_n$ in \eqref{PF3} is obtained. 
However, it seems more natural to introduce the order $p$ at the very 
beginning to make discussions with distinguishing manifestly the algebra 
and its representation. Thus, we have defined the parafermion algebra by 
\eqref{PF1}--\eqref{PF3}.
\nopagebreak\par
The Fock representation of PFA$_p$ is characterized by the unique vacuum 
$\rvac$ which satisfies
\begin{eqnarray}
\noalign{\vskip-15pt}
&& a_n \rvac = 0, \ \ \quad n=1,\,2,\,\ldots \label{pfvac1}\\
&& a_m a_n^* \rvac = p\,\delta_{m,n} \rvac.           \label{pfvac2}
\end{eqnarray}
The Fock space for parafermion is spanned by $\rvac$ and of 
$a_{n_1}^*\cdots a_{n_k}^*\,\rvac$ $(k=1,\,2,\,\ldots)$.
The positivity of the inner product of this Fock space is due to the fact 
that $p$ is a positive integer. 
\par
It is possible to embed PFA$_p$ into $\bigotimes\limits^p\,$CAR. Here, it 
should be noted that any element of CAR$\,\otimes{}I\otimes\cdots\otimes{}I$ 
{\it commutes\/} with that of 
$I\otimes\,$CAR$\,\otimes{}I\otimes\cdots\otimes{}I$, and likewise for any 
other combinations. This kind of description of parastatistical algebra is 
introduced by Green\cite{Green}. 
Let $a_n^{(\alpha)}$ $(\alpha=1,\,\ldots,\,p\,;\ n=1,\,2,\,\ldots)$,
which are called Green components, be generators of
$\overbrace{I\otimes\cdots\otimes I}\limits^{\alpha-1}\otimes\,$CAR$\,
\otimes\overbrace{I\otimes\cdots\otimes I}\limits^{p-\alpha}$. 
Then, they satisfy the unconventional anticommutation relations as follows:
\begin{eqnarray}
&& \{ a_m^{(\alpha)}, \, (a_n^{(\alpha)})^* \} = \delta_{m,n}I, 
                                                      \label{green1}\\
&& \{ a_m^{(\alpha)}, \, a_n^{(\alpha)} \} = 0,       \label{green2}\\
&& [ a_m^{(\alpha)}, \, a_n^{(\beta)} ] 
   = [ a_m^{(\alpha)}, (a_n^{(\beta)})^* ]
   =0   \ \ \ \text{ for } \alpha \not= \beta.        \label{green3}
\end{eqnarray}
We denote the $\ast$-algebra generated by the Green components 
$\{ a_n^{(\alpha)}\, | \, \alpha=1,\,\ldots,\,p,\, n=1,\,2,\,\ldots\}$ 
by GCA$_p$ ({\it Green-component algebra of order $p$\/}) 
$(\cong\bigotimes\limits^p\,$CAR). 
Then, an embedding $\iota$ of PFA$_p$ into GCA$_p$ is given by a linear 
combination of the Green components as follows
\begin{eqnarray}
\noalign{\vskip-8pt}
&&
\begin{array}{c}
\iota : \mbox{PFA}_p \hookrightarrow \mbox{GCA}_p, \\[2pt]
\displaystyle\iota(a_n) \equiv \sum_{\alpha=1}^p a_n^{(\alpha)}.
\end{array}
        \label{greencomp}
\end{eqnarray}
It is straightforward to show that \eqref{greencomp} satisfies 
\eqref{PF1}--\eqref{PF3}. 
The vacuum satisfying \eqref{pfvac1} and \eqref{pfvac2} is given by the vacuum 
of Fock representation of GCA$_p$:
\begin{equation}
a_n^{(\alpha)} \, \rvac = 0, \ \ \alpha=1,\,2,\,\ldots,\,p,\, \  
n=1,\,2,\,\ldots.
\end{equation}
\vskip8pt
\subsection{Embedding of Parafermion Algebra}
We, now, consider a recursive construction of embedding of PFA$_p$ into 
$\calO_{2^p}$. Let $\bolda^{(\alpha)}\in\calO_{2^p}$, 
$\zeta_\alpha:\calO_{2^p} \to \calO_{2^p}$ be a linear mapping and 
$\varphi_\alpha$ a unital $\ast$-endomorphism of $\calO_{2^p}$ for 
$\alpha=1,\,2,\,\ldots,\,p$. Then, a set of $p$ triads
$RP_p=\{\, (\bolda^{(\alpha)};\,\zeta_\alpha,\,\varphi_\alpha) \, | \, 
\alpha=1,\,\ldots,\,p\,\}$ is called a {\it recursive parafermion 
system of oder $p$\/} RPFS$_p$, if it satisfies the following conditions:
\begin{eqnarray}
\noalign{\vskip-5pt}
\noalign{$\phantom{ii}\mbox{i) seed condition: } $\vskip5pt}
&&\qquad
 (\bolda^{(\alpha)})^2 = 0, \quad 
 \{ \bolda^{(\alpha)},\,\bolda^{(\alpha)}{}^* \} = I, \label{rpfs1}\\
&&\qquad
 [ \bolda^{(\alpha)}, \, \bolda^{(\beta)} ] = 0, \quad
   [ \bolda^{(\alpha)}, \, \bolda^{(\beta)}{}^* ] = 0 \ \ \ 
    \text{ for } \alpha\not=\beta,                    \label{rpfs2}\\
\noalign{$\phantom{i}\mbox{ii) recursive condition: }$\vskip5pt\nopagebreak}
&&\qquad 
\{ \bolda^{(\alpha)},\, \zeta_\alpha(X) \} = 0,   \quad
   X \in \calO_{2^p},          \label{rpfs3}\\
&&\qquad
[ \bolda^{(\alpha)},\, \zeta_\beta(X) ] = 0 \ \ \ 
    \text{ for } \alpha\not=\beta,                    \label{rpfs4}\\
\noalign{$\mbox{iii) normalization condition: }$\vskip5pt}
&&\qquad
\zeta_\alpha(X)\zeta_\alpha(Y)=\varphi_\alpha(XY),  \quad
   X, \, Y \in \calO_{2^p},   \label{rpfs5}\\
&&\qquad
[\zeta_\alpha(X), \, \zeta_\beta(Y)]=0 \ \ \ 
    \text{ if }\ [ X, \, Y ]=0, \ \  
  \text{for } ^\forall \alpha,\, \beta.               \label{rpfs6}
\end{eqnarray}
Then, an embedding $\varPhi_{RP_p}$ of GCA$_p$ into
$\calO_{2^p}$ is determined as
\begin{equation}
\begin{array}{c}
\varPhi_{RP_p} : \mbox{GCA}_p \hookrightarrow \calO_{2^p}, \\[8pt]
\varPhi_{RP_p}(a_n^{(\alpha)})\equiv \zeta_\alpha^{n-1}(\bolda^{(\alpha)}),
\quad \alpha=1,\,2,\,\ldots,\,p, \ n=1,\,2,\,\ldots,
\end{array}                                           \label{rpfs7}
\end{equation}
which satisfy \eqref{green1}--\eqref{green3} as shown by straightforward 
calculation. 
Therefore, an embedding 
of PFA$_p$ into $\calO_{2^p}$ 
is obtained by restricting \eqref{rpfs7} to PFA$_p\subset\,$GCA$_p$
as follows
\begin{eqnarray}
&&
(\varPhi_{RP_p}\circ\iota)(a_n) = \sum_{\alpha=1}^p 
    \zeta_\alpha^{n-1}(\bolda^{(\alpha)}), \quad
 n=1,\,2,\,\ldots.                                    \label{rpfs8}
\end{eqnarray}
\par
First, let us consider the case $p=2$. 
We apply \eqref{gRFS1}--\eqref{gRFS4} to $\calO_4$.
According to the way of dividing indices $\{1,\,2,\,3,\,4\}$ into two parts 
in \eqref{gRFS1}, we obtain some $\bolda$'s, among which we can easily 
find a pair of $\bolda$'s commuting with each other. 
Here, we give an example, what we call the {\it standard RPFS$_2$\/}, 
$SRP_2=\{\,(\bolda^{(\alpha)};\,\zeta_\alpha,\,\varphi_\alpha) \, | \, 
\alpha=1,\,2 \, \}$:
\begin{eqnarray}
&& \bolda^{(1)} \equiv s_1 s_2^* + s_3 s_4^*,         \label{a1}\\
&& \zeta_1(X) \equiv s_1 X s_1^* - s_2 X s_2^* 
                  + s_3 X s_3^* - s_4 X s_4^*,        \label{phi1}\\
&& \bolda^{(2)} \equiv s_1 s_3^* + s_2 s_4^*,         \label{a2}\\
&& \zeta_2(X) \equiv s_1 X s_1^* + s_2 X s_2^* 
                  - s_3 X s_3^* - s_4 X s_4^*,        \label{phi2}\\
&& \varphi_1(X)=\varphi_2(X)=\rho(X)\equiv 
                    s_1 X s_1^* + s_2 X s_2^* 
                  + s_3 X s_3^* + s_4 X s_4^*.        \label{rho}
\end{eqnarray}
As in the case of the standard RFS$_2$, it is shown that 
$\varPhi_{SRP_2}(\hbox{GCA}_2)$ is identical with $\calO_4^{U(1)}$. 
\par
One should note that \eqref{a1} and \eqref{a2} can be obtained from 
$(\bolda_1, \, \bolda_2)$ of the standard RFS$_2$ defined by \eqref{O_4-a} 
and \eqref{O_4-a2} through the Klein transformation as follows:
\begin{eqnarray}
&& \bolda^{(1)} = \bolda_1, \\
&& \bolda^{(2)} = ( I - 2 \bolda_1^* \bolda_1 ) \, \bolda_2 
                = \big[\exp(i\pi \bolda_1^* \bolda_1)\big]\, \bolda_2.
\end{eqnarray}
Furthermore, $\zeta_\alpha, \ \alpha=1,\,2$ defined by \eqref{phi1} and 
\eqref{phi2} satisfy the following
\begin{eqnarray}
&& 
\zeta_1^{n-1} (X) 
= \Big[\exp\Big(i\pi \sum_{k=1}^{n-1} 
  \varPhi_{SR_2}(a_{2k})^* \varPhi_{SR_2}(a_{2k})\Big)\Big]\,\zeta^{n-1}(X),\\
&& 
\zeta_2^{n-1} (X) 
= \Big[\exp\Big(i\pi \sum_{k=1}^{n-1} 
  \varPhi_{SR_2}(a_{2k-1})^* \varPhi_{SR_2}(a_{2k-1})\Big)\Big]\,
  \zeta^{n-1}(X), 
\end{eqnarray}
where $\zeta$ is defined by \eqref{O_4-phi} and use has been made of an 
identity $\zeta(XYZ)=\zeta(X)\zeta(Y)\zeta(Z)$.  Thus, the Green components
$\{ \varPhi_{SRP_2}(a_n^{(\alpha)}) \}$ are rewritten in terms of
generators of CAR of the standard RFS$_2$ $\{ \varPhi_{SR_2}(a_n) \}$ 
\eqref{UHF-RFS2} through the Klein transformation defined by
\abovedisplayskip=10pt\belowdisplayskip=10pt
\begin{eqnarray}
&& 
\varPhi_{SRP_2}(a_n^{(1)}) \!=\!
 \begin{cases}
   \varPhi_{SR_2}(a_1) & \mbox{for $n=1$,} \\[5pt]
   \displaystyle 
   \Big[\!\exp\Big(i\pi\sum_{k=1}^{n-1} 
   \varPhi_{SR_2}(a_{2k})^* \varPhi_{SR_2}(a_{2k})\Big)\!\Big]\, 
   \varPhi_{SR_2}(a_{2n-1}) 
   & \mbox{for $n\geqq2$,}
 \end{cases} \qquad \\
&& 
\varPhi_{SRP_2}(a_n^{(2)}) \!=\!
  \Big[\!\exp\Big(i\pi\sum_{k=1}^{n} 
   \varPhi_{SR_2}(a_{2k-1})^* \varPhi_{SR_2}(a_{2k-1})\Big)\!\Big]\, 
   \varPhi_{SR_2}(a_{2n}),
\end{eqnarray}
which gives an automorphism of $\calO_4^{U(1)}$.
It should be noted that it is impossible to lift the above Klein 
transformation to a unital $\ast$-endomorphism of $\calO_4$.
\par
For the case of generic $p$, we construct the standard RPFS$_p$ 
$SRP_p=\{(\bolda^{(\alpha)}; \, \zeta_\alpha,\,\phi_\alpha) \, | \,
\alpha=1,\,\ldots,\,p\}$ as follows:
Let $s_i$'s be generators of $\calO_{2^p}$.
For an arbitrarily fixed $\alpha$, each term $s_i s_j^*$ in 
$\bolda^{(\alpha)}$ requires that $\bolda^{(\beta)}$ $(\beta\not=\alpha)$ 
should involve either $s_i s_{k_\beta}^* + s_j s_{\ell_\beta}^*$ 
or $s_{k_\beta} s_i^* + s_{\ell_\beta} s_j^*$ with some $k_\beta$ 
and $\ell_\beta$, and then $\bolda^{(\alpha)}$ is conversely required to 
involve $s_{k_\beta}s_{\ell_\beta}^*$ for all $\beta$. 
In this way, by setting 
$\bolda^{(1)}\equiv\sum\limits_{k=1}^{2^{p-1}}s_{2k-1}s_{2k}^*$, 
the explicit expressions for 
$(\bolda^{(\alpha)},\, \zeta_\alpha,\,\varphi_\alpha), \  
\alpha=1,\,2,\,\ldots,\,p$, are given by
\begin{eqnarray}
&&\bolda^{(\alpha)}
   =\sum_{k=1}^{2^{p-\alpha}}
    \sum_{\ell=1}^{2^{\alpha-1}}
    s_{2^\alpha(k-1)+\ell}s_{2^{\alpha-1}(2k-1)+\ell}^*,  
                                                      \label{rpfs_p-1}\\
&&\zeta_\alpha(X)
   =\sum_{i=1}^{2^p}(-1)^{\left[\frac{i-1}{2^{\alpha-1}}\right]}
    s_i X s_i^*,                                      \label{rpfs_p-2}\\
&&\varphi_\alpha(X)
     =\rho(X)\equiv\sum_{i=1}^{2^p}s_i X s_i^*,       \label{rpfs_p-3}   
\end{eqnarray}
where $[x]$ denotes the largest integer not greater than $x$.
It is shown that $(\varPhi_{SRP_p}\circ\iota)(\hbox{PFA}_p)$ is a proper 
subset of $\calO_{2^p}^{U(1)}=\varPhi_{SRP_p}(\hbox{GCA}_p)$.
In the same ways as in RPFS$_2$, it is possible to rewrite the Green 
components of RPFS$_p$ $\{ \varPhi_{SRP_p}(a_n^{(\alpha)})\}$ in terms of 
generators of the standard RFS$_p$ $\{ \varPhi_{SR_p}(a_n) \}$ using the 
Klein transformation. 
\vskip30pt
%
%
%
\section{Discussion}
In the present paper, we have introduced the most fundamental aspect of the 
recursive fermion system (RFS) in the Cuntz algebra $\calO_{2}$ and its 
generalization (RFS$_p$) in $\calO_{2^p}$, and apply it to construct the 
recursive parafermion system of order $p$ (RPFS$_p$).
As explicit examples, we have presented the standard RFS and the standard 
RFS$_p$ which give embeddings of CAR onto the $U(1)$-invariant subalgebra 
$O_2^{U(1)}\!\cong\,$UHF$_2$ and $\calO_{2^p}^{U(1)}\!\cong\,$UHF$_{2^p}$, 
respectively.
Although they are the simplest and most important examples, we can construct 
other ones explicitly using unital $\ast$-endomorphisms of the Cuntz algebra. 
As for the canonical endomorphism, it commutes with the $U(1)$ action, hence 
its restriction to the standard RFS yields another RFS which gives an 
embedding onto a proper subset of $\calO_2^{U(1)}$.  
Since, however, a generic unital \hbox{$\ast$-endomorphism} does {\it not\/} 
necessarily commute with the $U(1)$ action, there exists a RFS $R$ such 
that $\varPhi_R(\mbox{CAR})\nsubseteqq\calO_2^{U(1)}$.
For example, the following unital $\ast$-endomorphisms $\varphi_i$ 
$(i=1,\,2)$ of $\calO_2$ do not commute with the $U(1)$ action:
\begin{eqnarray}
&&\varphi_1(s_1) = s_{1;\,1} + s_{21;\,2}, \quad \varphi_1(s_2) = s_{22}, 
                                                      \label{inhom-endo}\\
&&\varphi_2(s_1) = s_{2;\,1} + s_{12;\,2}, \quad \varphi_2(s_2) = s_{11}.
\end{eqnarray}
\par
In our construction of embeddings of the CAR algebra into the Cuntz algebra, 
the indices of fermion operators $\{ \varPhi_R(a_n) \}$\ $(n=1,2,\ldots\,)$ 
denote the ordering of generations yielded recursively by the recursive map 
$\zeta\,: \,\varPhi_R(a_{n-1}) \mapsto \varPhi_R(a_n)$\ $(n=1,2,\ldots\,)$.
However, in the physical point of view, they should be interpreted as modes 
which distinguish momentum of particle or other physical degrees of freedom. 
The relation between the recursive map $\zeta$ and such physical meaning of 
indices of fermion operators is still an open problem. 
\par
Description of the CAR algebra of fermions in terms of the Cuntz algebra is 
{\it not\/} just a rewriting of well-known things. In our succeeding 
paper\cite{AK2}, we will discuss on various applications of RFS and RFS$_p$ 
by restricting the properties of the Cuntz algebra. 
In the conventional viewpoint in the $C^*$-algebra, the structure of the Cuntz 
algebra has been studied through that of the UHF algebra (UHF$_2\cong\,$CAR), 
since the latter is believed to be understood enough.  
However, inverting the way of consideration, quite a new viewpoint will 
open and show us various novel structures of CAR.
For example, by restricting endomorphisms and automorphisms of the Cuntz 
algebra to RFS or RFS$_p$,  we can explicitly construct nontrivial 
endomorphisms of CAR as mentioned above and automorphisms which are expressed 
in terms of {\it nonlinear\/} transformations.\cite{AK3} 
The Bogoliubov transformations, which are expressed in terms of linear 
transformations of the annihilation/creation operators, are no more than 
almost trivial examples in this context. 
We will also find that infinite branching of vacuum of fermions occurs 
through a certain type of endomorphisms including \eqref{inhom-endo}. 
\par
In gauge theories and quantum gravity, fermions called Faddeev-Popov (FP) 
ghosts play quite an important role at the fundamental level of theory.
In our another succeeding paper,\cite{AK4} we will consider recursive 
construction of FP ghost algebra in string theory.  Since it is possible 
to formulate the FP ghosts only on the basis of the indefinite-metric state 
vector space, we need to generalize the Cuntz algebra in such a way that it 
acts on the indefinite-metric vector space.  The resultant Cuntz-like 
algebra is called the {\it pseudo Cuntz algebra.}  We will construct two 
embeddings of the FP ghost algebra in string theory into the pseudo Cuntz 
algebra and discuss on restricted representations. The special attention will 
be payed for the zero-mode FP ghost operators.
\vskip30pt
%
%
%
%
%
%
%

\end{document}